\documentclass[twocolumn, english, aps, nobalancelastpage, superscriptaddress, raggedbottom, amsmath, amssymb, amsfont, eprint]{revtex4-1}
\usepackage[colorlinks]{hyperref} 
\hypersetup{
  colorlinks   = true, 
  urlcolor     = blue, 
  linkcolor    = blue, 
  citecolor   = red 
}

\usepackage[latin9]{inputenc}
\usepackage{xcolor}
\usepackage{float}
\usepackage{graphicx}
\usepackage{babel}
\usepackage{braket}
\usepackage{siunitx}
\usepackage{ulem}
\usepackage{multirow}
\usepackage[caption=false]{subfig}

\newcommand*{\balancecolsandclearpage}{
  \clearpage
  \twocolumngrid
}

\begin{document}
\title{Preferred Interaction Ranges in Neutral-Atom Arrays in the Presence of Noise}

\author{David C. Spierings}
\thanks{These authors contributed equally to this work. Email: edwin@entanglednetworks.com or david@entanglednetworks.com}
\affiliation{Entangled Networks Ltd.\\Toronto, ON, M4R 1A2, Canada}
\affiliation{
 Centre for Quantum Information and Quantum Control, Department of Physics, University of Toronto, Toronto, Ontario, Canada.
}
\author{Edwin Tham}
\thanks{These authors contributed equally to this work. Email: edwin@entanglednetworks.com or david@entanglednetworks.com}
\affiliation{Entangled Networks Ltd.\\Toronto, ON, M4R 1A2, Canada}

\author{Aharon Brodutch}
\affiliation{Entangled Networks Ltd.\\Toronto, ON, M4R 1A2, Canada}
\author{Ilia Khait}
\affiliation{Entangled Networks Ltd.\\Toronto, ON, M4R 1A2, Canada}

\begin{abstract}
Successful execution of a quantum information processing (QIP) task on a quantum processing device depends on the availability of high-quality entangling gates.
Two important goals in the design and implementation of any entangling gate are low error rates and high connectivity.
The former minimizes unintended perturbations to the quantum state during application of that gate, while the latter maximizes the set of qubits that can interact directly without remapping the QIP task through intermediary qubits -- a step that can require many additional gates.
Unfortunately, these goals can sometimes conflict, necessitating a careful trade-off.
In this work, we study that trade-off in two-dimensional (2D) arrays of neutral atoms interacting through two-qubit gates mediated by the Rydberg blockade effect.
The connectivity associated with Rydberg mediated gates on a 2D array is limited by the strength of the Rydberg blockade shift, which decays with distance.
Whereas a common strategy to improving connectivity is to use Rydberg levels with larger dipole moments, doing so also leaves the atom more susceptible to electric field noise.
Here, we simulate the performance of various logical QIP operations under realistic noise sources and for a variety of Rydberg levels in order to evaluate the connectivity versus gate error trade-off.
We find that under many noise regimes, a preferred range of interaction emerges that best satisfies that trade-off.
While the exact optimum interaction range depends closely on the details of the atomic implementation, we present simple scaling arguments with broad applicability that should inform future hardware and compiler design choices.
\end{abstract}

\maketitle

\section{Introduction}
Quantum computing (QC) promises a host of exciting applications~\cite{brassard2002qaa,grover1996,farhi2014quantum,peruzzo2014vqe,shor1999polynomial}. Amidst these promises, the field of quantum information processing (QIP) has seen rapid advances in real-world implementations. This rapid progress has brought about realisations of quantum computers on a plethora of physical systems~\cite{pino2021demonstration,bartolucci2021fusion,linke2017experimental,gambetta2017building}.

\begin{figure}[h]
\begin{centering}
\includegraphics[width=0.9\columnwidth]{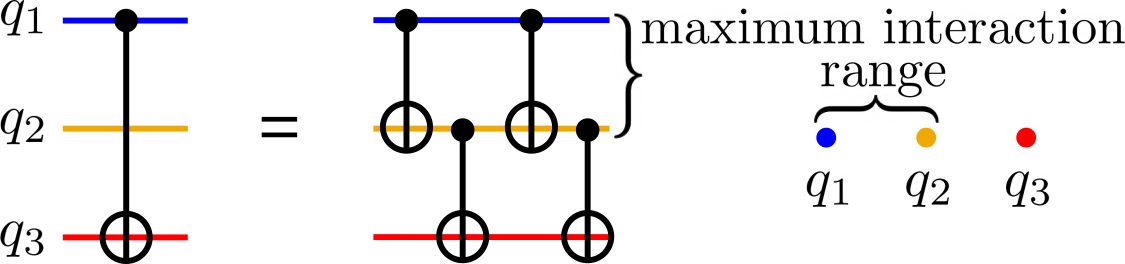}
\par\end{centering}
\caption{Example of a controlled-X operation between distant qubits (left), synthesized from a sequence of short-ranged operations (middle). Here, the physically permissible operations are determined by the qubit layout (right). \label{fig:synthesis}}
\end{figure}

In any quantum computer, a fundamental challenge is to engineer high-quality non-local gates that can entangle/disentangle states of disparate physical qubits -- a prerequisite for a quantum speed-up over classical computers. Physical qubit implementations tend to have a limited set of neighbors with which direct interactions are viable for high-quality entangling operations. Such limits on ``connectivity" are typically dictated by the physical mechanism that enables multi-qubit interactions.
Since a physical qubit often must interact with numerous others when performing a desired logical operation in QIP tasks, whenever an interaction is infeasible given the connectivity constraints of a QC platform, the desired logical operation must be ``synthesized'' during compilation from the set of permissible physical ones, often involving intermediary qubits~\cite{mosca2020,amy2018,gheorghiu2020}.
A simple example is shown in Fig.~\ref{fig:synthesis}, where a desired logical operation (in this case a controlled-X or CX) between qubits $q_1$ and $q_3$ is resynthesized using only CX operations between nearest-neighbor qubits (i.e.~$q_1\rightarrow q_2$ or $q_2\rightarrow q_3$).
Resynthesized operations that conform to restrictive hardware constraints tend to result in longer sequences of gates and therefore incur greater cost in runtime and noise. Thus, hardware architectures with greater connectivity can avoid the pitfalls of large overheads added at compile time.

In most QC platforms, the underlying hardware connectivity is determined by manufacturing and physical constraints. For instance, a single crystal of trapped ions affords a mechanism due to their Coulomb interaction that can couple any pair of ions, resulting in an all-to-all topology within the crystal. On the other hand, solid-state devices like superconducting chips tend to be arranged in topologies where every qubit is connected to a very limited number of neighbors. Understanding the trade-offs between increased connectivity versus other considerations (e.g.~noise, gate times, qubit count etc.) on any QC platform is important for the proper design of the quantum processing unit (QPU) itself as well as a compiler toolchain that can efficiently map QIP tasks onto it~\cite{murali2020}.

Arrays of cold, neutral atoms are a promising and unique QC platform as they supply rather flexible connectivity~\cite{Saffman_2010,Whitlock_2021}. Neutral atoms are held in place via optical traps that can provide reconfigurable geometries, both before and during a computation~\cite{Saffman_demo_2019,Lukin_2022,Saffman_demo_2022,Atomcomputing_2022}. In addition, entangling operations are supported by Rydberg interactions, whose range can be tuned by the choice of Rydberg levels, as well as through the tuning of external fields~\cite{Ryabtsev_2010,Ravets_2014}. Long-range connectivity in a neutral-atom array can, however, limit the number of operations permitted to occur at the same time (i.e.~in parallel). 
Recently, the implications of restricted gate parallelism in a neutral-atom array were analyzed, and a general reduction in overheads with increased Rydberg interaction range was nevertheless demonstrated~\cite{Chong_2021}.

In this work, we expand on that analysis by considering realistic noise models in a neutral-atom array. Specifically, we explore the trade-off between increased connectivity with the use of longer-range Rydberg interactions versus the increased susceptibility to noise resulting from correspondingly higher Rydberg levels. Our analysis suggests an optimum Rydberg interaction range
that minimizes overall error. We find that technical noise can be mitigated up to a point through the use of longer-range interactions that reduce total gate count in a compiled circuit. Beyond that point, further increase in Rydberg interaction range is impeded by noise that scales rapidly with the range of Rydberg interactions. This favored interaction range allows for the conceptualization of ``clusters'' of atoms packed within that range. Knowing the optimal cluster size \textit{a priori} is a useful fact that can aid the design of compilers, for example, that intelligently partition circuits so as to minimize inter-cluster operations that incur greater synthesis overheads.


In Sec.~\ref{sec:Model} we first outline a simple model for QC in neutral-atom arrays. Section \ref{subsec:PhysicalNoise} summarizes the dipole-dipole interaction that typically mediates entangling operations in neutral-atom arrays. Here we emphasize the role of noise sources that depend on the interaction range, as well as the significance of noise which is range independent. In Sec.~\ref{subsec:PhysicalComputingModel}, we define the physical computing model that we analyze, which is a two-dimensional neutral-atom array utilizing the Levine-Pichler gate for entangling operations~\cite{Levine_Pichler_2019}, and in Sec.~\ref{subsec:NumericalNoise} we define the numerical model that quantifies system performance. Section \ref{sec:numerical_results} presents numerical simulations and explicates simple scaling arguments that underpin our results. Section \ref{sec:Bench} illustrates our results in the context of more complicated circuits by benchmarking a variety of QIP tasks under the same computing and noise models. Finally, in Sec.~\ref{sec:conclusions}, we summarize our results, discuss their implications, and propose future directions of study.


\section{Computing Model\label{sec:Model}}

\subsection{Rydberg Atom Interactions and Sensitivity\label{subsec:PhysicalNoise}}

Multi-qubit interactions between neutral atoms trapped in arrays are mediated by the electric-dipole coupling between Rydberg states. These interactions are orders of magnitude larger than the short range interactions between groundstate neutral atoms due to the large radius of Rydberg states, which scales like $n^2$, where $n$ is the principal quantum number of the Rydberg state. Rydberg interactions can extend over many microns, enabling interactions between multiple neighbors in a neutral-atom array. The strength and position dependence of interactions is set by the exact atomic states used, which in general provides significant flexibility to tune the qubit connectivity in a neutral atom quantum computer.

One prominent example of Rydberg interactions, which we will focus on here, is the Rydberg blockade effect in which a single Rydberg excitation shifts the energies of the Rydberg states of nearby atoms out of resonance, preventing a second excitation for the affected atoms. Rydberg blockade is most commonly studied in the `long-range' limit, where resonant interactions between Rydberg states can be ignored and the dipole interaction Hamiltonian can be treated using second-order perturbation theory, yielding
\begin{equation}
    \hat{H}_\text{vdW}=-\frac{C_6^{(a,b)}}{r^6}\ket{a,b}\bra{a,b},
    \label{eq:H_vdW}
\end{equation}
where $C_6^{(a,b)}$ is the van der Waals coefficient determining the strength of interactions between two atoms in the Rydberg states $a$ and $b$. The volume of atoms that experience a significant energy shift is characterized by the Rydberg blockade radius, $r_\text{b}\equiv (|C_6^{(a,b)}|/\Omega)^{(1/6)}$, defined as the region within which $\langle H_\text{vdW}\rangle$ is significant compared to the excitation Rabi frequency $\Omega$. More precisely, the Rydberg-Rydberg interaction is determined by a sum over contributions from all dipole-coupled pair states with forms akin to Eq.~\ref{eq:H_vdW}. Often experimentally, and here for simplicity, the two Rydberg states are chosen to be the same, both share the same principle quantum number $n$, and both have angular momentum of zero (i.e.~s-orbitals). Hence, $a=b=n$ and we write the van der Waals coefficient simply as $C_6^{(n)}$. In this case, the Rydberg-Rydberg interactions for alkali atoms are generally isotropic, repulsive, and have van der Waals coefficients that scale like $C_6^{(n)}\sim n^{12}$~\cite{Saffman_2008}.

The long-range interactions of Rydberg states owe their utility to their strong electric-dipole couplings. This also makes Rydberg states extremely sensitivity to background electric fields. The energy shift, $\delta E$, of a Rydberg level due to a static electric field of magnitude $E$ is given by
\begin{equation}
    \delta E = \frac{1}{2}\alpha E^2,\label{eq:field_noise}
\end{equation}
where $\alpha$ is the scalar polarizability of the Rydberg state~\cite{Saffman_2005}. In general, the polarizability scales with the principal quantum number of the Rydberg level as $\alpha\sim n^7$.
This strong scaling of the sensitivity to stray electric fields makes large Rydberg atoms fragile objects. Though this sensitivity does not provide a fundamental limit to the utility of Rydberg states for entangling operations, as one could imagine working in the absence of background field fluctuations, it is worthwhile to understand the restrictions such noise will impose in practical scenarios. For instance, one may intuitively aim to increase connectivity in a neutral atom quantum computer by working with the longest achievable interaction ranges. Due to the limited parallelism necessary in common quantum algorithms, this has in fact already been demonstrated to be a gratuitous approach~\cite{Chong_2021}. Yet, in the presence of noise it may also be a disadvantageous one. Rydberg levels with principal quantum numbers of many hundreds have been accessed experimentally~\cite{Frey_1996,Tannian_2000}, but two-qubit gates using Rydberg levels in the range of roughly $n\sim50-150$ are being explored for near-term quantum computation in part due to the presence of stray electric fields \cite{Levine_2018,Levine_Pichler_2019}. State-of-the-art neutral atom quantum computers experience spurious electric fields of magnitude around $10\si{mV\per cm}$~\cite{Lukin_2022,Saffman_demo_2022}.

Still, modern neutral atom quantum computers using relatively low-lying Rydberg levels are mostly limited by a variety of technical obstacles, rather than by noise induced by stray electric fields. Specifically, the fidelity of two-qubit gates is hampered by errors that are not fundamental to the entangling protocols being implemented. For instance, in alkali atoms the main mechanism for Rydberg addressing is via a two-photon transition that boosts the effective two-photon Rabi frequency by a relatively small detuning to an intermediate state. Scattering a spontaneously emitted photon from this intermediate state during a two-qubit gate can be a sizeable contribution to the infidelity of entangling operations. This source of noise depends most significantly on the single-photon detuning from the intermediate state, and is roughly independent of the choice of Rydberg level. Though there are other technical sources of noise (e.g.~nonzero atom temperature, pointing instability of lasers, etc.) that do not depend on the Rydberg level used for long-range interactions, for the moment we focus only on this scattering mechanism for simplicity. Typical scattering times in recent experiments are $50-100\si{\micro\second}$~\cite{Lukin_2022,Saffman_demo_2022}.


\subsection{Rydberg Computing Model \label{subsec:PhysicalComputingModel}}

In order to investigate the interplay of the different noise sources discussed in Sec.~\ref{subsec:PhysicalNoise} and their consequences for a given computation, we analyze a topology and two-qubit gate scheme that are relevant for state-of-the-art systems, but also simple enough to gain intuition. 
In brief, we evaluate the performance of a square, two-dimensional (2-D) grid of neutral atoms in which the entangling operations are implemented using the Levine-Pichler (LP) protocol~\cite{Levine_Pichler_2019}. In the following we summarize the physics relevant to our analysis. 

\subsubsection{Levine-Pichler Gate\label{subsubsec:LP_gate}}
The Levine-Pichler gate implements a controlled-phase gate between two qubits encoded in long-lived states of two atoms. Like other schemes, the LP gate works in the Rydberg blockade regime, but is novel in that it does not require independent manipulation of the two qubits and uses continuous Rabi pulses that can minimize the time spent in the Rydberg state~\cite{Jandura_2022}. In the most simple implementation, the LP gate is a two pulse sequence characterized by three parameters: the duration, $\tau,$ and detuning, $\Delta$, which are the same for each pulse, and the relative phase $\xi$ between the pulses. The pulses are simultaneously applied to both atoms and only couple one of the qubit states, say $\ket{1}$, to a Rydberg state $\ket{r}$ with Rabi frequency $\Omega$, while the other qubit state, $\ket{0}$, is uninvolved. In the fully blockaded limit (i.e.~the approximation that double occupancy of the Rydberg state is completely prohibited), the two-qubit states $\ket{01}$ and $\ket{10}$ evolve as two-level systems with excited states $\ket{0r}$ and $\ket{r0}$, respectively. The state $\ket{11}$ also evolves as a two-level system, but with the excited state $\ket{W_{1r}}=\left(\ket{1r}+\ket{r1}\right)/\sqrt{2}$ and with Rabi frequency $\sqrt{2}\Omega$ due to the collective enhancement~\cite{Lukin_2000_enhance,Saffman_2015_enhance}. The two-qubit state $\ket{00}$ is of course unaffected by the pulse sequence. The difference in Rabi frequency driving the evolution of the $\ket{01}$ (and $\ket{10}$) versus $\ket{11}$ systems means that they follow different trajectories on the Bloch spheres representing their respective two-level systems, and thus generally acquire distinct geometric phases after the pulses. When set appropriately, the three gate parameters, $\tau, \Delta, \xi,$ are sufficient to ensure not just that the $\ket{01}$, $\ket{10}$, and $\ket{11}$ states perform a complete, detuned Rabi oscillation, but also that they acquire the necessary phase difference so as to be maximally entangling:
\begin{align}
\ket{00}&\rightarrow\ket{00}\nonumber\\
\ket{01}&\rightarrow\ket{01}e^{i\phi}\nonumber\\
\ket{10}&\rightarrow\ket{10}e^{i\phi}\nonumber\\
\ket{11}&\rightarrow\ket{11}e^{2\phi-\pi}\label{eq:LPaction2}
\end{align}
which is equivalent to a controlled-Z (CZ) gate up to a single qubit phase rotation of $\phi$. Specifically, in the fully blockaded limit, the parameters are set such that
\begin{equation}
    \tau=\frac{2\pi}{\tilde{\Omega}(0)},
    \label{eq:tau}
\end{equation}
where $\tilde{\Omega}(\delta)\equiv\sqrt{2\Omega^2+(\Delta+\delta)^2}$ is the generalized Rabi frequency for the evolution of the $\ket{11}$ state. Thus, without noise ($\delta=0$), the $\ket{11}$ state performs a complete Rabi oscillation during each pulse of the LP gate. Additionally, the phase difference between the pulses is determined from~\cite{Levine_Pichler_2019}
\begin{equation}
    e^{-i\xi}=\frac{iy\sin{\big(\frac{1}{2}s\sqrt{y^2+1}\big)}-\sqrt{y^2+1}\cos{\big(\frac{1}{2}s\sqrt{y^2+1}\big)}}{iy\sin{\big(\frac{1}{2}s\sqrt{y^2+1}\big)}+\sqrt{y^2+1}\cos{\big(\frac{1}{2}s\sqrt{y^2+1}\big)}},
    \label{eq:xi}
\end{equation}
where $y=\Delta/\Omega$ and $s=\Omega\tau$, so that the $\ket{01}$ and $\ket{10}$ states evolve back to their original states after two partial rotations, and $\Delta$ is set to impart the phase shifts in Eq.~\ref{eq:LPaction2}. Here, given an experimentally achievable Rabi frequency $\Omega,$ we numerically solve the two-level equations of motion to determine $\Delta$, and calculate $\tau$ and $\xi$ according to Eqs.~\ref{eq:tau} and \ref{eq:xi}.

The susceptibility of the Rydberg state's energy to background electric field fluctuations, described in Sec.~\ref{subsec:PhysicalNoise}, impacts the output state of the LP gate through the detuning. On a given operation of the LP gate, the true detuning (i.e.~$\Delta + \delta$, for $\delta\neq 0$) experienced by the two atoms undergoing the gate may differ from the intended one (i.e.~$\Delta$), which is used to set the parameters of the LP gate. The consequence is an error in the phase imparted by the phase gate, as well as lingering population in the Rydberg state $\ket{r}$ due to improper rotations during the Rabi oscillation.
We discuss modeling effects of \textit{physical} errors discussed here as \textit{logical} ones in Sec.~\ref{subsec:NumericalNoise}.

\subsubsection{Connectivity and Range of Rydberg Interactions\label{subsubsec:connectivity}} Connectivity among qubits within a QPU is an important consideration in QPU design since a low-connectivity device must compensate by \textit{synthesizing} long-range operations from a sequence of operations involving only directly-coupled qubits~\cite{mosca2020,gheorghiu2020,sabre2019,amy2018}. Synthesized operations necessarily involve more physical steps than direct logical gates, and suffer from correspondingly longer runtimes and more severe compounded gate errors.

The connectivity in a neutral atom quantum computer is determined by the comparison of two length scales: the separation between atoms and the range of Rydberg interactions. Here we consider architectures in which each atom represents one physical qubit, and the atoms are arranged in a 2-D, square lattice with unit distance $a$ that is fixed throughout a given computation. In experiments, the minimum separation between atoms on the lattice is often determined by practical considerations, such as the resolution of the imaging system used to read out qubit states, or the wavelength of the optical trapping potential when the atoms are confined in the standing wave produced by interfering beams. With these constraints, minimum separations on the order of a $1-20\si{\micro\m}$ are typical. For simplicity, we consider only static 2-D lattices, and omit from our analysis movable optical tweezer traps that allow a lattice to be reconfigured mid-computation~\cite{Lukin_2022}.

\begin{figure}[h]
\begin{centering}
\includegraphics[width=8.5cm]{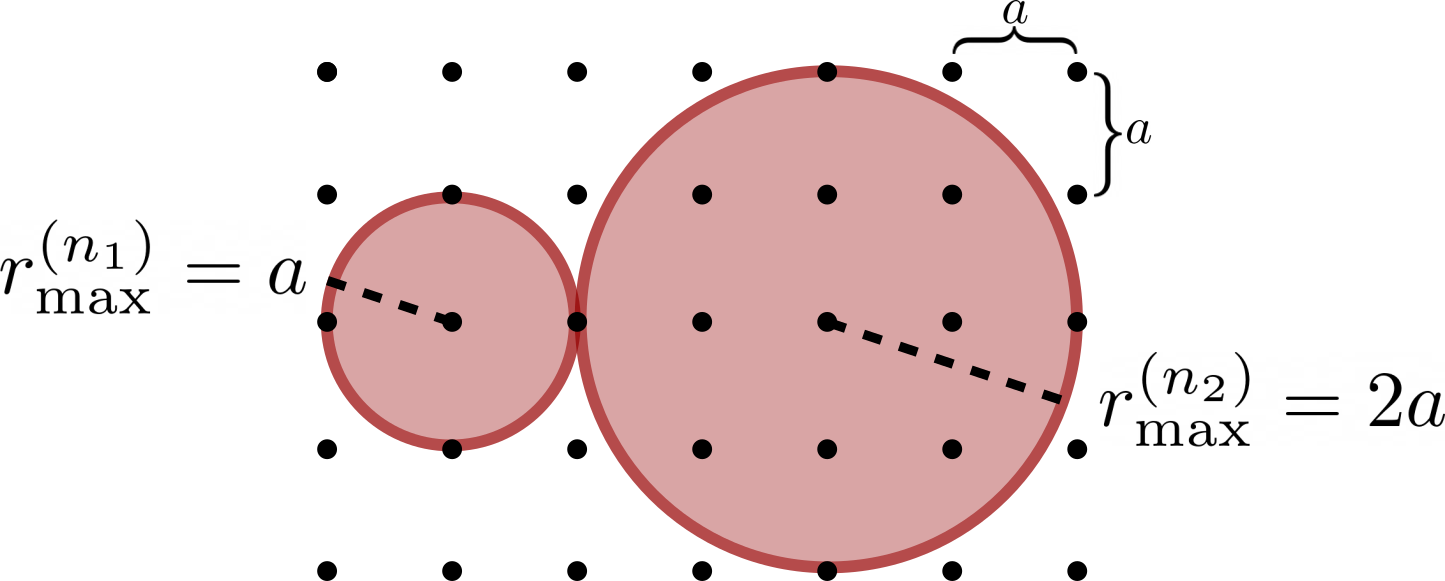}
\par\end{centering}
\caption{Illustration of Rydberg blockade mediated operations on a two-dimensional neutral-atom array with lattice spacing $a$. Two interaction ranges of radii $r_\text{max}^{(n_1)}$ and $r_\text{max}^{(n_2)}$ are achieved by Rydberg levels with principal quantum numbers of $n_1$ and $n_2$. Longer interaction ranges allow qubits to interact directly over longer distances without resorting to intermediate operations, but incur the cost of preventing more atoms from being able to compute in parallel as well as being more susceptible to noise.\label{fig:array_cluster}}
\end{figure}

In the limit of sufficiently large separations, the range of Rydberg interactions between two atoms is set by the van der Waals interaction. As mentioned in Sec.~\ref{subsec:PhysicalNoise}, one metric to define its range is the Rydberg blockade radius. For the purposes of a two-qubit gate, however, the atom-atom separation at which an entangling gate can be performed with high fidelity is determined by the details of the entangling protocol. For a gate that works in the Rydberg blockade regime, e.g.~LP gates, the approximation that $\langle\hat{H}_\text{vdW}\rangle\gg \Omega,\Delta$ must be satisfied.
One way to ensure this is to choose Rydberg levels such that the energy shift of the doubly occupied Rydberg state is $\langle\hat{H}_\text{vdW}\rangle\sim k\Omega,k\Delta$, where $k>1$. In this way, the constant $k$ determines the necessary blockade shift above which two-qubit gates can be performed reliably. This sets the maximum interaction range of a Rydberg blockade mediated gate according to
\begin{equation}
    r_\text{max}^{(n)}=\bigg(\frac{C_6^{(n)}}{k\Omega}\bigg)^{1/6}.
\label{eq:Threshold}\end{equation}
Under this condition for high-fidelity two-qubit gates, we establish that only atoms separated by a distance less than $r_\text{max}^{(n)}$ can be entangled through a single application of the LP gate, as portrayed in Fig.~\ref{fig:array_cluster}. Logical operations between atoms separated by more than $r_\text{max}^{(n)}$ on the other hand must be logically synthesized by applying a sequence of one and two-qubit gates, as was exemplified in Fig.~\ref{fig:synthesis}.

It is worth noting a few additional details. While the equations that set the gate parameters of the LP gate, outlined in Sec.~\ref{subsubsec:LP_gate}, rely on the validity of the fully blockade approximation (i.e.~$\langle H_\text{vdW}\rangle\gg\Omega,\Delta$) the principal of the LP gate does not. For finite interaction potentials comparable to $\Omega$ and $\Delta$, it is still possible to find trajectories that bring all two-qubit states back to their origin, and supply the desired phase shifts. These trajectories become more complicated for weaker blockade shifts, and thus in practice harder to implement. For the simulations summarized here, we make the simplification that all LP gates are performed with a large enough blockade shift that use of the equations for the gate parameters, Eq.~\ref{eq:tau} and \ref{eq:xi}, is justified. Lastly, Rydberg interactions set a minimum separation within which the blockade effect is no longer an appropriate description of the atom-atom interactions and also when Rydberg-ground state interactions in the lattice could become problematic~\cite{Saffman_2008}. Currently, the practical constraints of neutral-atom arrays, which inform our choice of lattice spacing, $a$, are such that experiments do not yet approach this maximum packing limit.

\subsection{Numerical Model \label{subsec:NumericalNoise}}

In order to understand the implications of a faulty implementation of the LP gate on quantum information processing (QIP) tasks, we model its effects on quantum states that have information-processing significance -- states that, jointly, we will refer to as the ``computational Hilbert space'' and denote by ${\cal H}_{C}$. An LP gate defect can occur when gate parameters $\tau$, $\Delta$, and $\xi$ are poorly-calibrated or are modulated by noise, and manifests as an under- or over-rotation (i.e. $\tilde\Omega(\delta)\tau\neq 2\pi$) during the Rabi oscillation described in Sec.~\ref{subsec:PhysicalComputingModel}. This transforms the basis states of ${\cal H}_{C}$ as follows:
\begin{align}
\left|00\right> & \to\left|00\right>\nonumber \\
\left|01\right> & \rightarrow\sqrt{p_{01}}e^{i\phi}\left|01\right>+\sqrt{1-p_{01}^{2}}\left|0r\right>\nonumber \\
\left|10\right> & \to\sqrt{p_{10}}e^{i\phi}\left|10\right>+\sqrt{1-p_{10}^{2}}\left|r0\right>\nonumber \\
\left|11\right> & \to\sqrt{p_{11}}e^{i\varphi}\left|11\right>+\sqrt{1-p_{11}^{2}}\left|W_{1r}\right>\label{eq:LLgate-noise-damping}.
\end{align}

Since the dynamical phases accrued during the LP gate depend on $\tilde\Omega$ and $\tau$, errors in those propagate to $\phi$ and $\varphi$ and can result in a coherent phase error (in the case of a calibration error) or a dephasing error (if gate parameters fluctuate).
However, a more pernicious manifestation of the transformations in Eq.~\ref{eq:LLgate-noise-damping} is loss. Specifically, loss of the portion of the wavefunction that is not in ${\cal H}_{C}$ is assumed to be unrecoverable, so it is omitted and the wavefunction renormalized. In such a situation, on any given shot (attempt to execute a QIP task) there is a finite probability that we might not find results corresponding to a valid statevector in ${\cal H}_{C}$ that corresponds to an ``answer'' to our computational task at all. Further, even if a loss of a qubit from ${\cal H}_{C}$ were not to occur on a given instance of an LP gate, the mere possibility of a loss perturbs the statevector within ${\cal H}_{C}$.

Under this error mode, an input two-qubit state $\left|\psi\right>=\sum_{a,b}c_{a,b}\left|a,b\right>$ evolves under a (noisy) LP gate like
\begin{equation}
\left|\psi\right>\to\frac{\sum_{a,b}c_{a,b}\sqrt{p_{ab}}e^{i\theta_{ab}}\left|a,b\right>}{\sqrt{1-P_{\text{loss}}}}\label{eq:LLgate-noise-loss},
\end{equation}
with probability $1-P_{\text{loss}}$. On the other hand, with probability $P_{\text{loss}}$, the state is lost (it is no longer in ${\cal H}_{C}$). Here, 
\begin{align}
P_{\text{loss}} & =1-\sum_{ab}p_{ab}\left|c_{ab}\right|^{2}\nonumber \\
\theta_{ab} & =\begin{cases}
0 & \text{if }a=b=0\\
\phi & \text{if }a\oplus b=1\\
\varphi & \text{if }a=b=1
\end{cases}\label{eq:Loss-params}
\end{align}
and we implicitly take $p_{00}=1$. In this error mode, in addition to a ``back-action'' whereby an intended output state is degraded via modulation with $p_{ab}$ (as in Eq.~\ref{eq:LLgate-noise-loss}), one also accrues a probability of loss that compounds with repeated application of the LP gate that can severely diminish the likelihood of obtaining \textit{any} answer to a computation.

Since the loss process is not completely-positive and trace-preserving (CPTP), simulating the process numerically with common quantum simulator frameworks requires a workaround that is fairly straightforward (we used IBM's QISkit Aer simulator~\cite{qiskit2022}). We first introduce an ancilla register, and consider the following unitary operator:
\begin{align*}
U_{\text{loss}}= & \left|00\right>_{\psi}\left<00\right|\otimes\mathbb{I}\\
 & +\left|01\right>_{\psi}\left<01\right|\otimes{\cal R}_{y}(\omega^{01})\\
 & +\left|10\right>_{\psi}\left<10\right|\otimes{\cal R}_{y}(\omega^{10})\\
 & +\left|11\right>_{\psi}\left<11\right|\otimes{\cal R}_{y}(\omega^{11})
\end{align*}
Here, the right-most register addresses the ancilla. Now suppose we apply $U_{\text{loss}}$ to the two-qubit state discussed above alongside an ancillary qubit prepared in $\left|0\right>$, i.e.~we compute $\ket{\psi^{'}}=U_{\text{loss}}\left(\left|\psi\right>\otimes\left|0\right>\right)$. If we set $\omega_{ij}=\arccos\left(\sqrt{p_{ij}}\right)$ then subsequent measurement of the ancillary qubit emulates the loss process described in Eqs.~\ref{eq:LLgate-noise-loss}~and~\ref{eq:Loss-params}. If the ancillary qubit is measured in the state $0$, then $\ket{\psi^{'}}$ is left in the state in Eq.~\ref{eq:LLgate-noise-loss}. On the other hand, if the ancillary qubit is found in state $1$, then a loss has occurred. Importantly, by tallying the fraction of shots for which all ancillary qubits (one for each instance of the LP gate) were found in state $0$, we can estimate the overall survival probability (i.e. $1-P_{\text{loss}}$).

We use the Alkali Rydberg Calculator (ARC) version 3.0 to determine realistic parameters for our numerics~\cite{ARC_3.0}. We calculate the pair-state interaction potentials and polarizabilities for $^{87}$Rb atoms in Rydberg levels $\ket{nS,m_j=1/2}$~\footnote{For clarity, $nS$ stands for $n$-th level and s-orbital (i.e.~zero orbital angular momentum).}, with principal quantum numbers in the range $n=50-110$. For the interaction potential of a given Rydberg level, we extract the van der Waals coefficient and find the maximum interaction radius as described in Sec.~\ref{subsubsec:connectivity}. Along with a fixed lattice spacing of $a=4\mu\text{m}$, this determines the connectivity/topology of the neutral atom QPU. We model background electric field noise of magnitude $10\si{mV\per cm}$. Inputting this field fluctuation into Eq.~\ref{eq:field_noise} along with the Rydberg level dependent polarizability, gives an error on the detuning of the LP gate of magnitude $\delta\equiv\delta E/\hbar$, from which we estimate $p_{ij}$ (see Sec.~\ref{sec:scaling_args}).

\section{Numerical Methods and Results}\label{sec:numerical_results}

\subsection{Single logical operation (CZ)\label{subsec:Single-logical-operation}}

In order to study the trade-off between interaction range and gate error in its simplest manifestation, we computed the effective overall error incurred when a \textit{logical} controlled-$Z$ operation is performed on a two-qubit computational state, as described in Sec.~\ref{sec:Model}, between pairs of qubits separated by various distances on a square lattice, using a variety of interaction ranges (and their corresponding principle quantum numbers). Specifically, given qubits arranged on a $10 \times 10$ lattice, we compute the effective error in performing a controlled-$Z$ operation between a qubit, $q_{1}$, residing on one corner of the lattice, at coordinate $(0,0)$, and some other qubit, $q_{2}$, at coordinate $(i,j)$. If the maximum Rydberg interaction distance ($r_{max}$) being modeled is shorter than the distance separating the two qubits, then we resort to logical synthesis methods that prescribe the insertion of additional physically permissible operations to effect the desired controlled-$Z$ gate. For the purposes of this section, we consider a one-way swap-based synthesis, wherein qubit $q_{1}$ is swapped (using physically permissible operations) with a sequence of consecutive neighbors until it is brought within proximity of $q_{2}$, whereupon the desired controlled-$Z$ gate is performed~\cite{ibmqxcomp2019}.

Note that for any given set of permissible physical interactions and a synthesis method, there tends to be multiple valid sequences of physical operations that all yield the correct logical gate (in this case a controlled-$Z$). An illustrative example is the following: supposing only nearest-neighbor interactions and swap-based synthesis, any route that connects the two qubits targeted by the desired controlled-$Z$ gate, traversing edges of the square grid, is a valid instantiation of that gate. We now briefly pose the problem of finding the ``best-possible'' instantiation under the given constraints as graph problem with known solutions.

Suppose we are permitted to use a maximum Rydberg interaction range of $r_{\text{max}}$. Because we had stipulated that physical qubits lie in a regular square grid of spacing $a$, this implies a set of integer 2-tuples:
\[
{\cal K}=\left\{ (k_{x,i},k_{y,i})\,\Big|\,k_{x,i},k_{y,i}\in\mathbb{Z},\,a\sqrt{k_{x,i}^{2}+k_{y,i}^{2}}\leq r_{\text{max}}\right\}. 
\]
Here, each 2-tuple is the coordinate of a ``neighbor'' that is accessible to a qubit via direct interaction (i.e.~no logical synthesis) that is physically permitted. Now further suppose we define a function $n(r)$ such that any LP gate operating between qubits separated by distance $r$ must use Rydberg level with principle quantum number $n(r)$ (we discussed how this function is chosen in Sec.~\ref{subsec:PhysicalNoise}). For each $n(r)$ (for $a\leq r\leq r_{\text{max}}$) we also associate a cost equal to the loss probability $P_{\text{loss}}(n(r))$ described in Sec.~\ref{subsec:NumericalNoise}.

We can now construct a graph $G(V,{\cal E},\omega)$ with nodes $V$ consisting of all physical qubits in the lattice and edges ${\cal E}$ consisting of all permissible physical interactions on that lattice. More precisely, a node $V_{a,b}$ situated at position $(a,b)$ within the grid has an edge to $V_{a+k_{x},b+k_{y}}$ for all $(k_{x},k_{y})\in{\cal K}$. Further, that edge has weight $\omega(a,b;k_{x},k_{y})=\log\left[P_{\text{loss}}\left(n(r)\right)\right]$, with $r=a\sqrt{k_{x}^{2}+k_{y}^{2}}$ -- so that traversal of that edge incurs a cost equivalent to the log-loss probability for a LP gate that uses Rydberg level with principle quantum number $n(r)$. Having constructed $G$, a controlled-$Z$ gate between $q_{1}$ and $q_{2}$ can be synthesized as a \textsl{route} on $G$ between nodes corresponding to those qubits. Specifically, we might imagine synthesizing a sequence of swaps along edges of $G$ starting from the $q_{1}$ node and ending at a \textit{neighbor} of the $q_{2}$ node. This is followed with a controlled-$Z$ between that neighbor (where the logical state held by $q_{1}$ now resides) and $q_{2}$. Notably, since we had explicitly constructed $G$ so that its edges all correspond to physically valid operations, this route between $q_{1}$ and $q_{2}$ implies a sequence of interaction radii ($r_{1}$, $r_{2}$, ... , $r_{m}$) that govern the LP gates that are performed along the way. This allows us to write an aggregated loss probability for the entire sequence of synthesized operations: 
\begin{align}
p_{\text{loss}}^{\text{Overall}}= & 1-\left[1-p_{\text{loss}}^{\text{CZ}}\left(n(r_{m})\right)\right]\nonumber \\
 & \times\prod_{k=1}^{m-1}\left[1-p_{\text{loss}}^{\text{Swap}}\left(n(r_{k})\right)\right]\nonumber \\
\log\left(1-p_{\text{loss}}^{\text{Overall}}\right)= & \log\left[1-p_{\text{loss}}^{\text{CZ}}\left(n(r_{m})\right)\right]\nonumber \\
 & +\sum_{k=1}^{m-1}\log\left[1-p_{\text{loss}}^{\text{Swap}}\left(n(r_{k})\right)\right]\label{eq:ErrorSwap}
\end{align}
In the latter line of Eq.~\ref{eq:ErrorSwap}, the sum is the aggregate log-likelihood of a loss incurred by the process of swapping qubit $q_{1}$ along a path that leaves it adjacent to $q_{2}$. Minimizing the sum in Eq.~\ref{eq:ErrorSwap} is equivalent to finding the shortest path between $q_{1}$ and neighbors of $q_{2}$ on $G$. Well-known algorithms exist that can efficiently compute the shortest weighted path between any pair of nodes. For our purposes, we resorted to an implementation of a Djikstra-based all-pair-shortest-path solver (APSP). We then proceeded to compute the overall log-loss (adding the single ``CZ'' term in Eq.~\ref{eq:ErrorSwap}) for each neighbor of $q_{2}$ (in all it has $\left|{\cal K}\right|$ neighbors), whereupon the best instantiation (lowest $P_{\text{loss}}^{\text{Overall}}$) of the target controlled-$Z$ gate can be chosen.

An important detail is that the precise value of $P_{\text{loss}}$, as defined (implicitly) in Eqs.~\ref{eq:LLgate-noise-loss}~and~\ref{eq:Loss-params}, depends on the quantum state on which the LP gate operates. In an optimistic extreme (say when the computational state is $\left|00\right>$) this loss is 0, whereas in a pessimistic extreme (when the computational state is $\left|11\right>$) it is maximal. In the numerical results we show here, $P_{\text{loss}}$ for various $n$ (and $r$) are computed as an average over 10,000 randomly initialised two-qubit states.

\subsubsection{Range-dependent noise\label{subsec:N-dependent-results}}
\begin{figure*}[ht]
\captionsetup[subfigure]{labelformat=empty}
\centering
\subfloat[\label{fig:ScaledLoss-GradedN}]{%
  \includegraphics[width=0.9\textwidth]{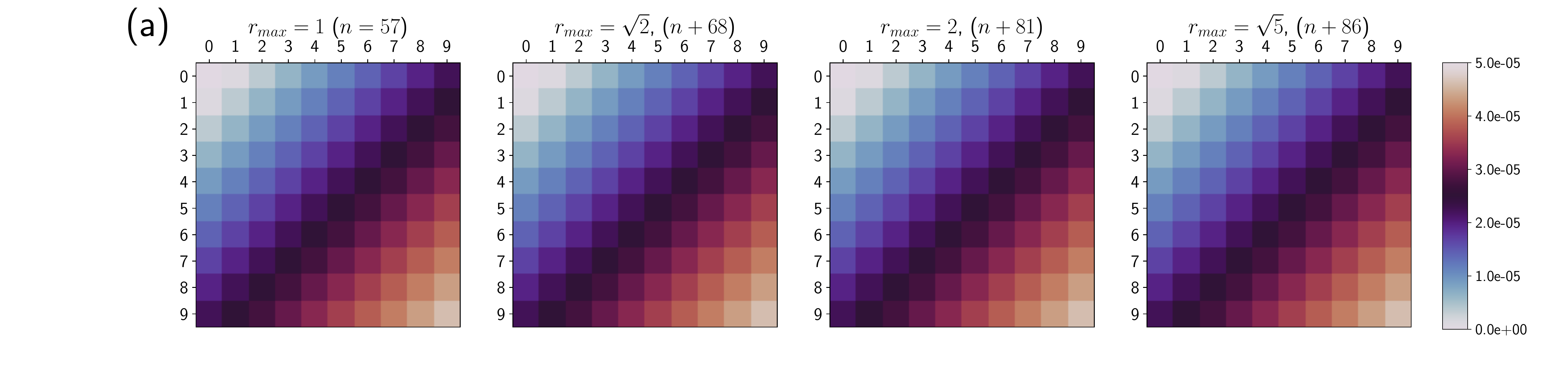}%
}
\vspace{-0.5cm}
\\
\subfloat[\label{fig:ScaledLoss-FixedN}]{%
  \includegraphics[width=0.9\textwidth]{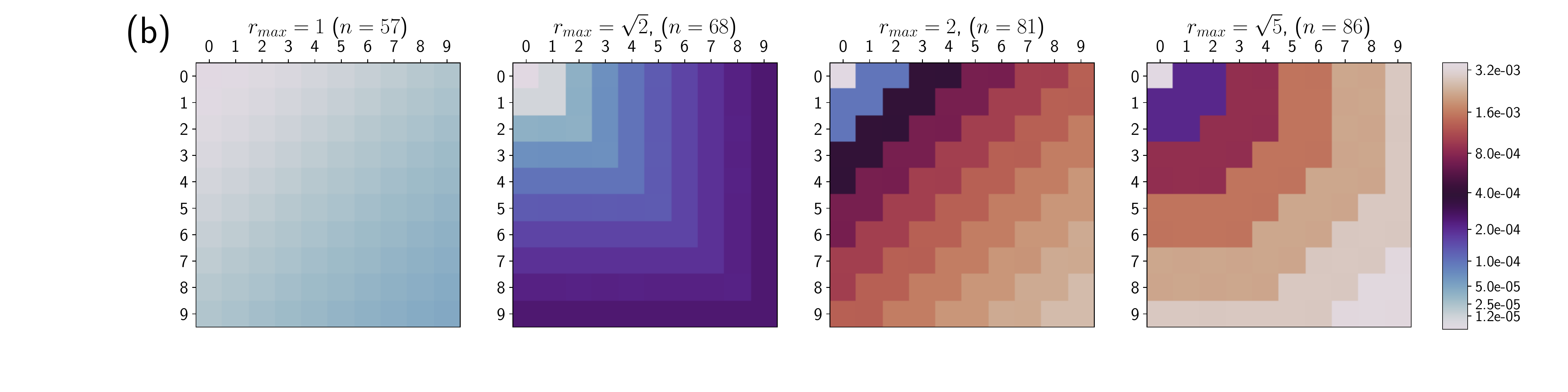}%
}
\vspace{-0.5cm}
\caption{Performance of a logical controlled-Z (CZ) gate realized on a 2-D neutral-atom lattice experiencing only electric field noise. In (a) and (b) a ``graded'' and ``fixed'' Rydberg level strategy is employed, respectively. Overall loss probabilities are shown as color-gradients for a CZ gate between location (0,0) and any other qubit on the $10\times10$ lattice. Physical gates are allowed to have a maximum interaction range (shown as $r_\text{max}$ above each grid) that varies for each plot, but every gate must use at least a Rydberg level that supports the given $r_\text{max}$. Gates requiring an interaction range greater than $r_\text{max}$ are synthesized as a sequence of swap operations plus a CZ.}
\end{figure*}

We start by probing the scenario where $P_{\text{loss}}^{\text{CZ}}$ in Eq.~\ref{eq:ErrorSwap} results \textit{only} from an error in $\delta$ due to stray fields described in Sec.~\ref{subsec:PhysicalNoise}. As seen in Eq.~\ref{eq:field_noise}, that error in $\delta$ and the corresponding loss $P_{\text{loss}}$ are strongly dependent on the principle quantum number $n$ and consequently on the Rydberg interaction range. A particularly important design choice that impacts the error probability is the choice of $n$ for a given $r$. Here, we consider two choices for $n(r)$: a ``graded'' as well as a ``fixed'' approach.

In a ``graded'' approach, we use the minimum viable principle quantum number for a desired interaction distance:
\[
n(r)=\min_{r\leq r_{\text{max}}^{(n)}}n
\]
That is, we pick the minimum $n$ such that the desired interaction range $r$ is still smaller than the maximum interaction radius defined in Eq.~\ref{eq:Threshold}. By contrast, in a ``fixed'' approach given a set of interaction ranges ${\cal R}$ we are interested in accessing, we use the minimum viable principle quantum number for that set:
\[
n(r\in{\cal R})=\min_{\max{\cal R}\leq r_{\text{max}}^{(n)}}n
\]
A graded approach is desirable for parallelism -- since only a single Rydberg-mediated gate can be executed within a blockaded region, being conservative by only enlarging the blockade region when necessary frees up the rest of the lattice for concurrent operations. On the other hand, supporting operations mediated by a large set of principle quantum numbers presents engineering and logistical challenges. Therefore, we also consider the case where a fixed principle quantum number is used, that is just sufficient to support the desired set of interaction ranges ${\cal R}$.

Figures \ref{fig:ScaledLoss-GradedN}~and~\ref{fig:ScaledLoss-FixedN} illustrate loss probabilities for a synthesized logical controlled-$Z$ operation between a qubit residing in the top-left corner of each $10\times10$ lattice, and a second qubit residing at various other sites in the same lattice. The ``graded'' approach to choosing $n$ is shown in Fig.~\ref{fig:ScaledLoss-GradedN} whereas the ``fixed'' approach is shown in Fig.~\ref{fig:ScaledLoss-FixedN}. Note that the color code is drawn on a logarithmic scale in order to accentuate minute variations in probability when only relatively small principal quantum numbers are used.

In both cases, observe that loss probabilities fail to improve with larger interaction radii. In other words, given only a range-dependent noise model where the noise scales rapidly with $n$, it is always preferable to use nearest-neighbor-only interactions to synthesize longer-range gates even if it results in a greater gate count. Further, it is noteworthy that whereas in the ``graded'' approach a compiler (or in this case an APSP algorithm) can choose \textit{not} to invoke operations involving high $n$'s, in the ``fixed'' approach by contrast \textit{any} design intent to use longer-range interactions necessitates the use of higher $n$'s for \textit{all} operations, leading to markedly worse outcomes in Fig.~\ref{fig:ScaledLoss-FixedN}.

\subsubsection{Inclusion of technical noise\label{subsec:N-independent-results}}
\begin{figure*}[ht]
\captionsetup[subfigure]{labelformat=empty}
\centering
\subfloat[\label{fig:FixedLoss-GradedN}]{%
  \includegraphics[width=0.9\textwidth]{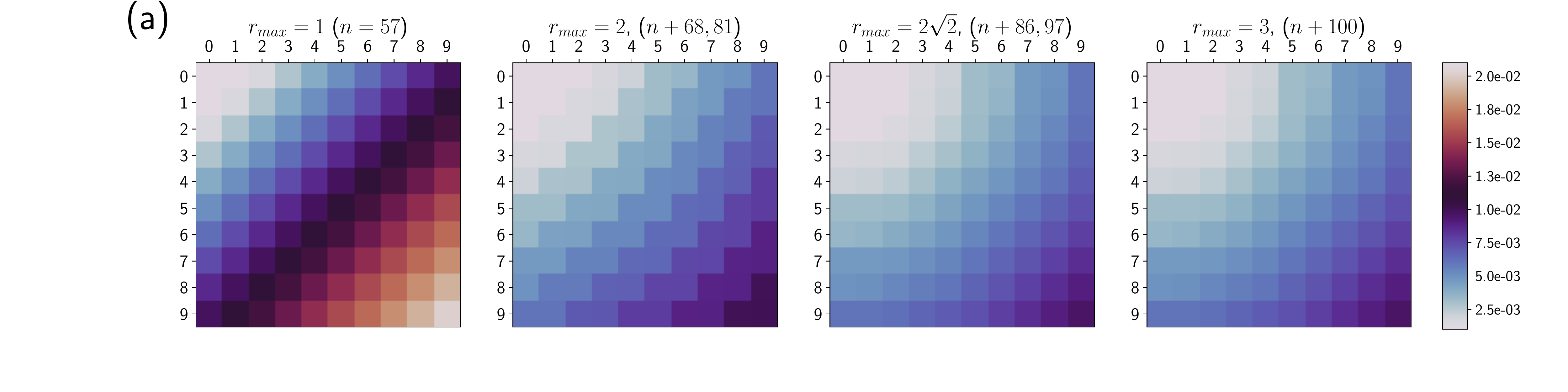}%
}
\vspace{-0.5cm}
\\
\subfloat[\label{fig:FixedLoss-FixedN}]{%
  \includegraphics[width=0.9\textwidth]{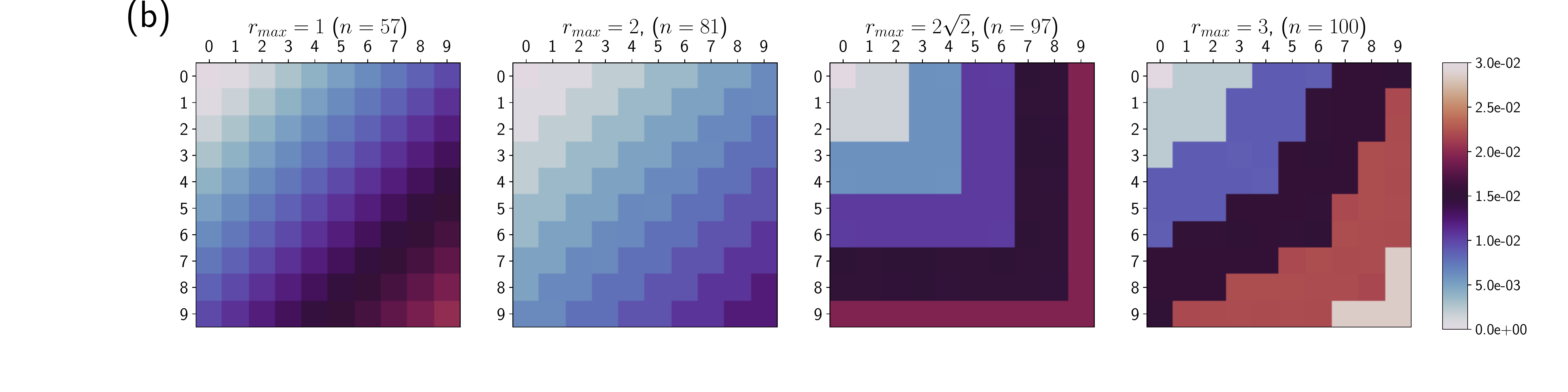}%
}
\vspace{-0.5cm}
\caption{Performance of a logical controlled-Z (CZ) gate realized on a 2-D neutral-atom lattice experiencing both electric field noise and errors due to photon scattering. In (a) and (b) a ``graded'' and ``fixed'' Rydberg level strategy is employed, respectively. Overall loss probabilities are shown as color-gradients for a CZ gate between location (0,0) and any other qubit on the $10\times10$ lattice. Physical gates are allowed to have a maximum interaction range (shown as $r_\text{max}$ above each grid) that varies for each plot, but every gate must use at least a Rydberg level that supports the given $r_\text{max}$. Gates requiring an interaction range greater than $r_\text{max}$ are synthesized as a sequence of swap operations plus a CZ.}
\end{figure*}

A model that better describes present-day implementations of Rydberg-mediated gates includes additional noise sources that do not scale with the principal quantum number of the Rydberg level. For low-lying Rydberg levels, overall noise is typically dominated by technical noise sources that are independent of $n$ and $r$, and the behaviour shown in Sec.~\ref{subsec:N-dependent-results} may cease to hold. While there are several sources of technical noise that contribute, we focus here only on scattering from an intermediate state used for the two-photon Rydberg transition. With the inclusion of this scattering (which we characterise with a scattering time $\tau_{\text{scat}}=50\mu\text{s}$), we repeat the numerical treatment above while including an $n$-independent noise in the form of a scattering probability $p_{\text{scat}}=\exp\left(-\tau/\tau_{\text{scat}}\right)$ for each LP gate, where $\tau$ is the gate time given in Eq.~\ref{eq:tau}.

We suppose that this scattering error is uncorrelated with the range-dependent loss, so that the overall survival probability is multiplicative: $1-p_{\text{loss}}^{\text{CZ}}\to\left(1-p_{\text{loss}}^{\text{CZ}}\right)\times\left(1-p_{\text{scat}}\right)$. Besides this extra multiplicative loss applied to each physical gate, all other aspects of the analysis proceeds as in Sec.~\ref{subsec:N-dependent-results}, yielding results shown in Figs.~\ref{fig:FixedLoss-GradedN} and~\ref{fig:FixedLoss-FixedN}.


Fig.~\ref{fig:FixedLoss-GradedN} shows loss probabilities in the case of ``graded'' principle quantum numbers, whereas Fig.~\ref{fig:FixedLoss-FixedN} shows the case of a ``fixed'' architecture. With purely range-dependent noise, the growth of noise with respect to the use of larger $n$ is such that any corresponding gain in connectivity -- and a resulting decrease in total number of physical operations -- is entirely negated. The upshot of the results in this section by contrast, is that when there is a large contribution of technical noise that is independent of $n$, there can indeed be a net benefit in using higher $n$'s in exchange for larger interaction ranges. This trend does not continue ad infinitum, however: once $n$ and $r$ becomes sufficiently large, the behaviour seen in Sec.~\ref{subsec:N-dependent-results} manifests yet again.

\subsection{Scaling arguments}\label{sec:scaling_args}
The qualitative observation of Secs.~\ref{subsec:N-dependent-results}~and~\ref{subsec:N-independent-results} that greater connectivity improves the performance of a synthesized logical operation only when range independent noise dominates gate errors can be understood from straightforward scaling arguments.

The loss probability, $p_{\text{loss}}$, increases with greater interaction range (i.e.~principal quantum number, $n$) due to over or under rotations from an improperly tuned LP gate, resulting from miscalibration of the detuning. If the detuning is incorrectly set so that $\Delta\to\Delta+\delta$, then the generalized Rabi frequency experienced by the $\ket{11}$ state will be $\tilde{\Omega}(\delta)=\sqrt{2\Omega^{2}+\left(\Delta+\delta\right)^{2}}$, but the gate time will still be $\tau=2\pi/\tilde{\Omega(0)}$, as discussed in Sec.~\ref{subsubsec:LP_gate}. In the presence of the noise $\delta$, there manifests an extraneous rotation on the Bloch sphere of magnitude $\tilde{\Omega}(\delta)\tau$.
Note that when $\delta=0$ then this corresponds to a $2\pi$ rotation, consistent with the ideal LP gate. 

The loss probability $p_{\text{loss}}$ is readily calculated from the population left behind in the $\ket{W_{1r}}$ state as given by detuned Rabi flopping in a two-level system,
\begin{align}
p_\text{loss}&=\frac{\Omega^2}{\tilde{\Omega}(\delta)^2}\sin^2{\bigg(\frac{1}{2}\tilde{\Omega}(\delta)\tau\bigg)}\\
 & \approx2\pi^{2}\Omega^{2}\frac{\delta^{2}\Delta^{2}}{\tilde{\Omega}(0)^{6}}\label{eq:ploss-scaling},
\end{align}
where we used Taylor expansions of $\sin^2$ around $\tilde{\Omega}(\delta)\tau\approx 2\pi$, and of $\tilde{\Omega}(\delta)$ around $\delta/\Delta\approx0$. In Sec.~\ref{subsec:PhysicalNoise}, it was noted that detuning error $\delta$ grows with increasing $n$ as $\delta\sim n^{7}$. Since $\delta$ is the only $n$-dependent term in Eq.~\ref{eq:ploss-scaling}, it follows that $p_{\text{loss}}\sim\delta^{2}$$\sim n^{14}$.

Now we turn our attention to the interaction range. Recall from Sec.~\ref{subsubsec:connectivity} that whether a principle quantum number can support a given interaction radius, is determined by comparing the Rydberg potential to a set multiple of the Rabi frequency. From Eq.~\ref{eq:Threshold} it follows that
\begin{align*}
-\frac{{\cal O}(n^{12})}{r^{6}} & \geq{\cal O}(1)\implies r\sim n^{2}
\end{align*}
The interaction range therefore grows much more slowly than $p_{\text{loss}}$ with respect to $n$. Put another way, $p_{\text{loss}}\sim r^{7}$. Taken together, this seemingly undergirds the observation in Figs.~\ref{fig:ScaledLoss-GradedN}~and~\ref{fig:ScaledLoss-FixedN}, where loss appears to worsen much faster than can be compensated by any gains in interaction range.

Finally, let us analyze behaviour in the presense of a technical noise. Suppose one wishes to perform a controlled-$Z$ gate between qubits that are connected through a chain of $D$ neighbours (in the formalism of Sec.~\ref{subsec:Single-logical-operation}, suppose $q_{1}$ and $q_{2}$ are connected along a route in $G$ with $D$ edges). The swap-based synthesis discussed above realizes the gate through $D-1$ swap operations and one controlled-$Z$ gate. If we further suppose all operations are mediated by the same Rydberg level, then Eq.~\ref{eq:ErrorSwap} reduces to:
\begin{align}
p_{\text{loss}}^{\text{Overall}} & =1-\left(1-p_{\text{loss}}^{\text{CZ}}\right)\left(1-p_{\text{loss}}^{\text{Swap}}\right)^{D-1}\nonumber \\
 & \approx1-\left(1-p_{\text{loss}}^{\text{CZ}}\right)^{3D-2}\label{eq:LossScalingOverall}
\end{align}
where we've used the fact that a swap is synthesizeable with three controlled-$Z$ operations. If we now consider an expansion of the interaction radius by a factor of $r$ and include $p_{\text{scat}}$, then Eq.~\ref{eq:LossScalingOverall} reads:
\begin{equation}
p_{\text{loss}}^{\text{Overall}}(r)=1-\left[\left(1-p_{\text{scat}}\right)\times\left(1-p_{\text{loss}}^{\text{CZ}}(r)\right)\right]^{\left(3D/r-2\right)}\label{eq:LossScalingD}.
\end{equation}
Supposing $p_{\text{scat}}=0$, by inspection note if $p_{\text{loss}}^{\text{CZ}}$ increases monotonically with $r$ (as we argued above, $p_{\text{loss}}\sim r^{7}$), then so too will $p_{\text{loss}}^{\text{Overall}}$. On the other hand, if $p_{\text{scat}}$ is finite, then monotonicity in $p_{\text{loss}}^{\text{Overall}}$ is broken.

\begin{figure}
\begin{centering}
\includegraphics[width=8.5cm]{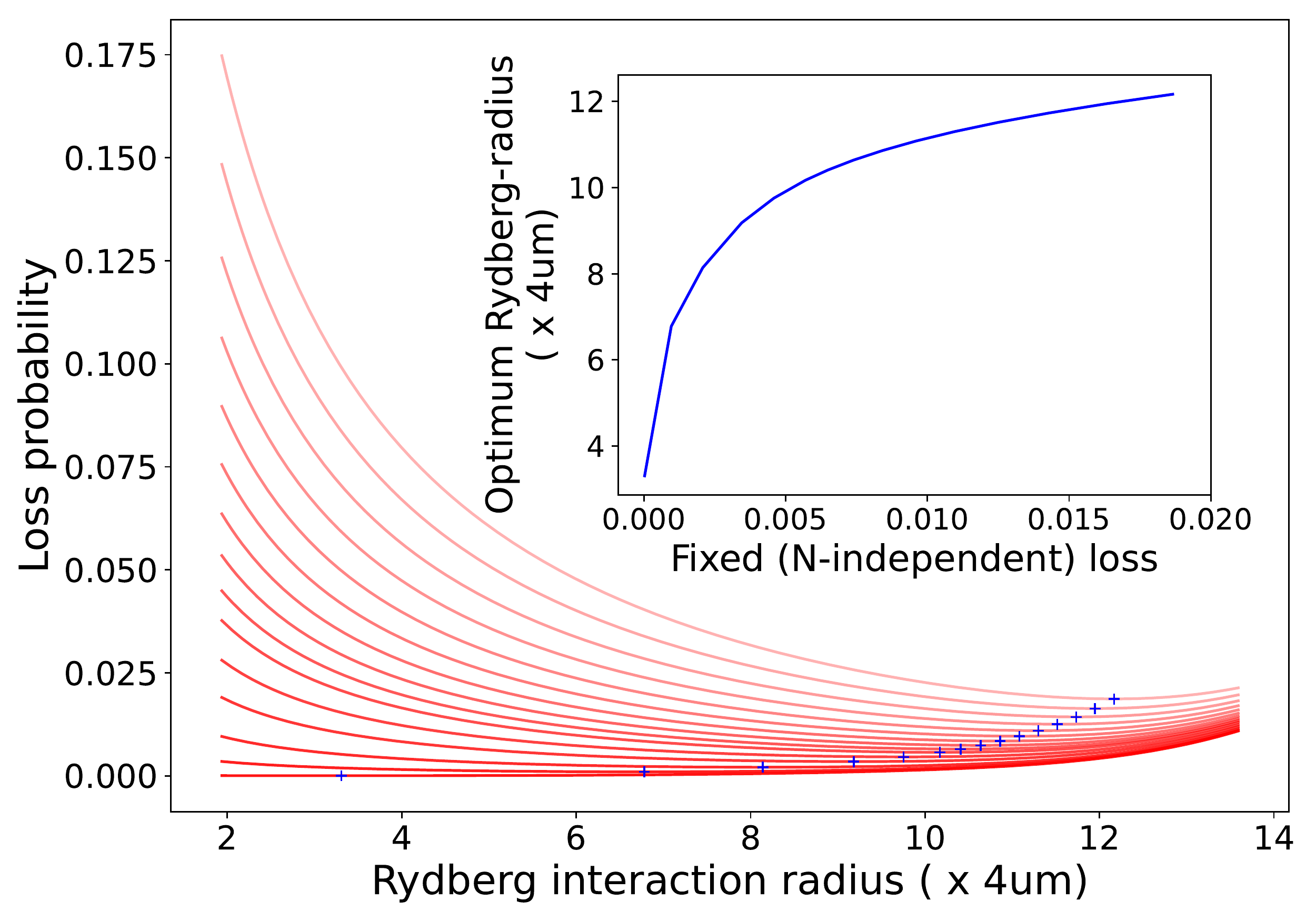}
\par\end{centering}
\caption{Plot of overall loss probability vs Rydberg interaction radius. Each red curve corresponds to a different loss contribution, $p_{\text{scat}}$, from technical noise sources characterized by varying scattering times, ranging from $\tau_{\text{scat}}=2\times10^{-5}s$ (lightest curve, highest fixed error rate) to $\tau_{\text{scat}}=0.1s$ (darkest, lowest fixed error rate). The blue crosses in the main plot indicate where minimas lie on each red curve. The inset plots the locus of those minimas against $p_{\text{scat}}$.\label{fig:LossPlot}}
\end{figure}

Fig.~\ref{fig:LossPlot} plots Eq.~\ref{eq:LossScalingD} for various choices of $p_{\text{scat}}$ and using the same $p_{\text{loss}}^{\text{CZ}}$ calculated for Secs.~\ref{subsec:N-dependent-results}~and~\ref{subsec:N-independent-results} for a ``graded'' $n(r)$. The figure illustrates the interplay of all the scaling behaviour discussed so far. Since the choice of $D$ has no bearing on salient features of Fig.~\ref{fig:LossPlot}, we arbitrarily set it to a convenient value of $D=14$. Note that optimum Rydberg interaction radii are more pronounced and sharply defined when the technical noise contribution, $p_{\text{scat}}$, is greater (lighter red curves). In the extreme when $p_{\text{scat}}\to0$ (darker red curve), the loss probability tends towards increasing monotonically with interaction radius.

We note that Fig.~\ref{fig:LossPlot} does not consider the discreteness of the lattice in which physical qubits lie, or the question of routing between points on that lattice, both of which were taken into account in Figs.~\ref{fig:ScaledLoss-GradedN}-\ref{fig:FixedLoss-FixedN}. Further, Eqs.~\ref{eq:LossScalingOverall}~and~\ref{eq:LossScalingD} ignore the fact that loss probabilities are, in fact, dependent on the specific quantum state at hand as discussed in Sec.~\ref{subsec:Single-logical-operation}. Nevertheless, we expect the essential conclusions from the simplified treatment in this section to hold even when those relatively more complicated considerations are taken into account.

\section{Circuit benchmark methods and results}\label{sec:Bench}
\begin{figure*}[ht]
\centering
\subfloat[\label{fig:BenchmarkNdep}]{%
  \includegraphics[width=0.95\columnwidth]{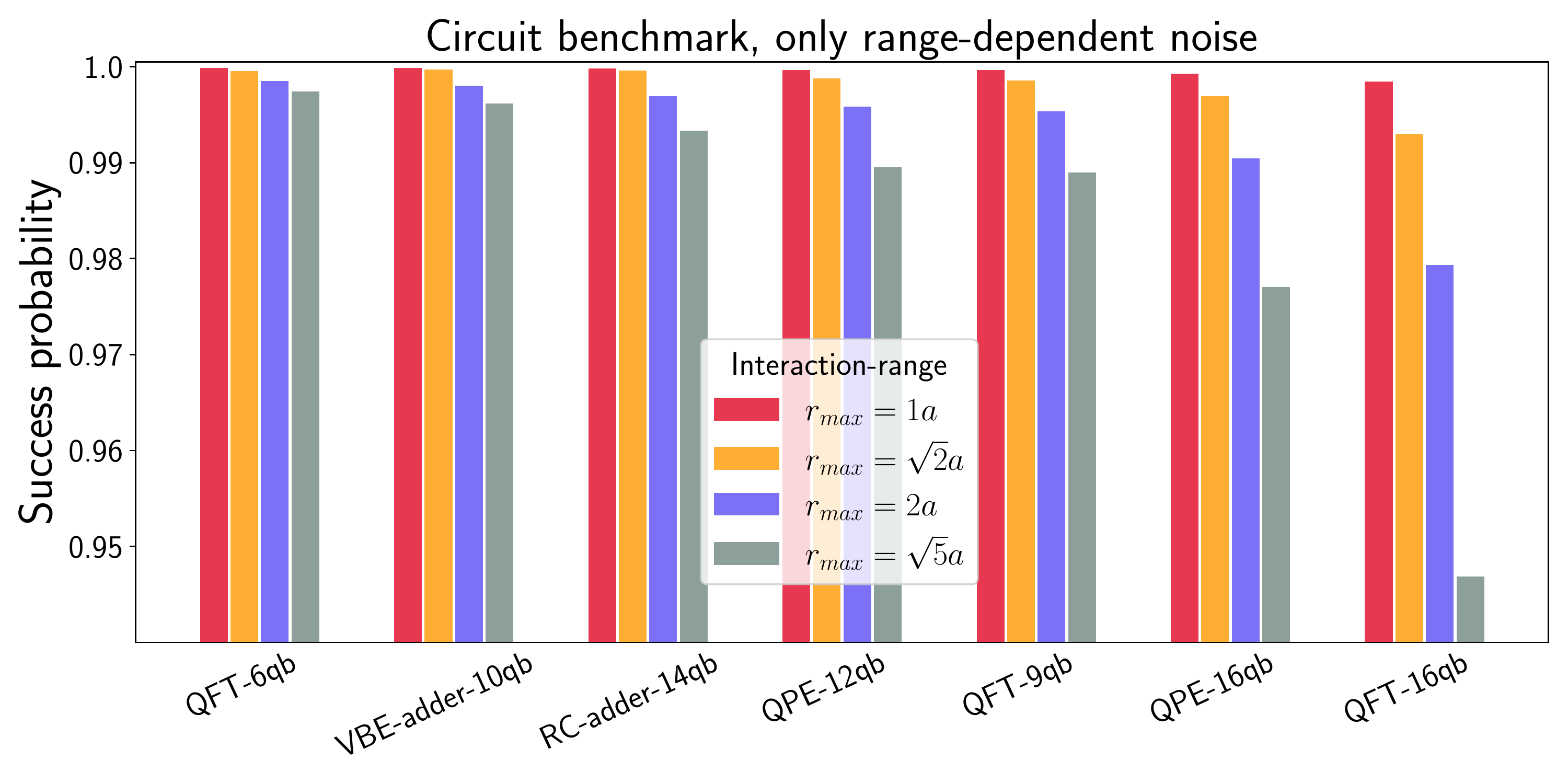}%
}
\subfloat[\label{fig:BechmarkNindep}]{%
  \includegraphics[width=0.95\columnwidth]{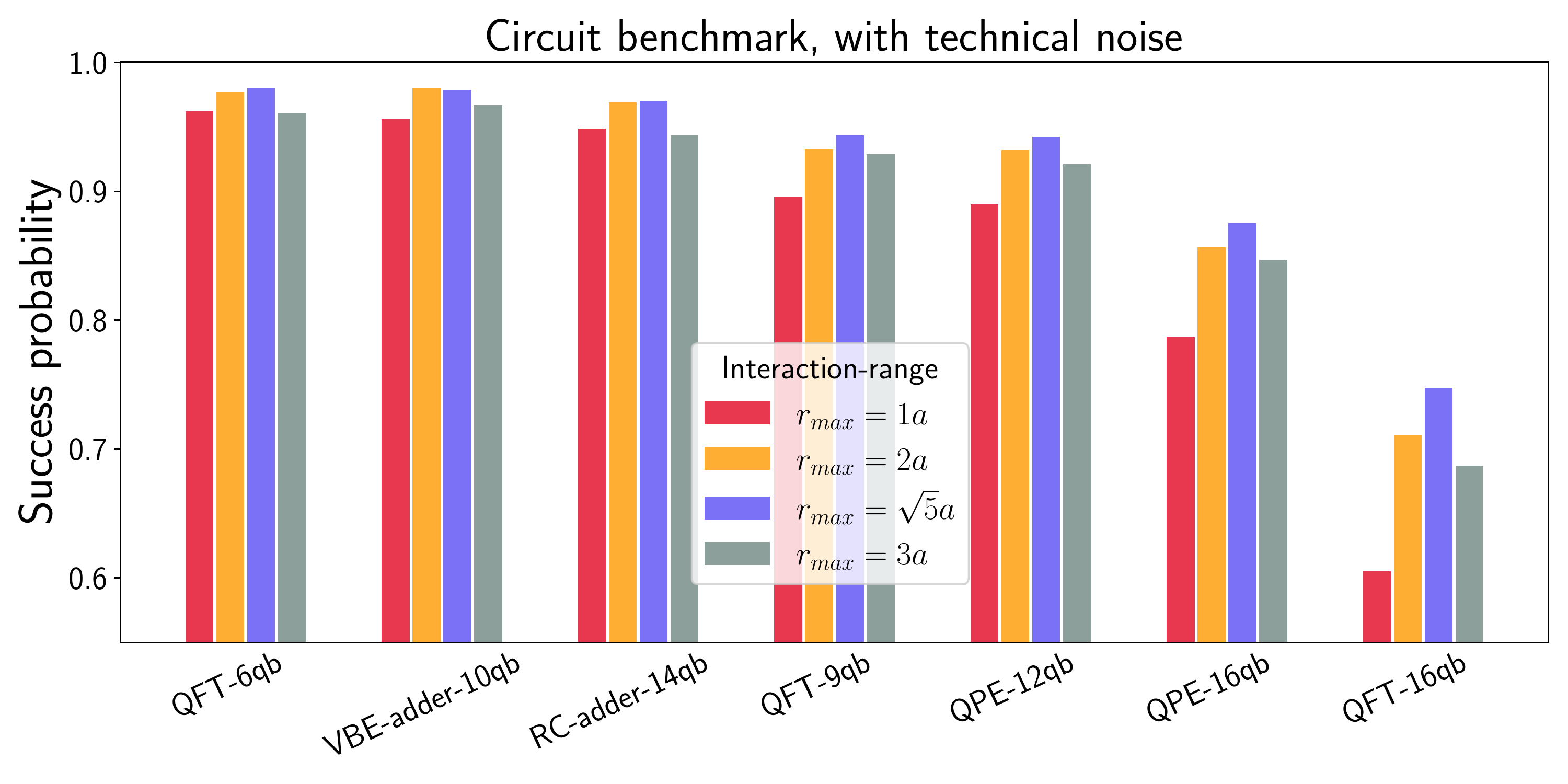}%
}
\caption{Benchmark of success probability given (a) only range-dependent noise that depends on principle quantum number $n$, and (b) both range-dependent noise and range-independent technical noise. Each bundle of bars corresponds to a different circuit, while the color code indicates the maximum interaction radius, $r_{\text{max}}$.}
\end{figure*}

Whereas the results in Sec.~\ref{sec:numerical_results} primarily concern the operation of a single logical controlled-$Z$ operation, in this section we verify that the same noise behaviour holds for more complicated circuits. To that end, we simulate a selection of benchmark circuits while applying the same noise models we have considered so far.

We select circuits from the Feynman and QED-C benchmark suites~\cite{feynman,amythesis2019,qedc,lubinski2021application}. The circuits are instances of the Quantum Fourier Transform (QFT), Quantum Phase Estimation (QPE), and adder arithmetic circuit families, spanning a range of widths (number of qubits) and depths (number of gates). Since the noise models we consider are not isotropic in the computational Hilbert space, performance of each circuit will be input dependent, favouring input data that leaves more rails in $\left|0\right>$ and incurring more noise when it leaves more rails in $\left|1\right>$. For each circuit family, we generate random inputs (e.g.~a phase for QPE, random integers for addition) and the corresponding quantum circuit.

Each circuit is passed through a compiler chain consisting of IBM's QISkit as well as a compiler written by Litteken et al., which is specific to neutral-atom arrays~\cite{nac}. The former performs basis reduction (i.e.~re-expressing the circuit purely in terms of controlled-$Z$ and arbitrary single-qubit gates). The latter is responsible for mapping logical qubits onto physical ones on the lattice, as well as routing and synthesis (essentially a more sophisticated analogue of the APSP algorithm discussed in Sec.~\ref{sec:numerical_results}) that accounts for Rydberg-blockade-related constraints~\cite{Chong_2021}. The resulting hardware-mapped circuit is then simulated, with an error imposed wherever a controlled-$Z$ gate is performed. Single-qubit gates are assumed to be ideal. Finally, by comparing the simulated output with the expected outcome given a random input, we compute the ``success probability'' as the probability that a correct answer is observed.

Figure \ref{fig:BenchmarkNdep} shows benchmark results in the case where only range-dependent noise is considered (it is analogous to Fig.~\ref{fig:ScaledLoss-GradedN}), while Fig.~\ref{fig:BechmarkNindep} adds a fixed technical noise that does not depend on interaction range (it is analogous to Fig.~\ref{fig:FixedLoss-GradedN}).
In both figures, each bar corresponds to an average over a set of thirty random input choices. Figure \ref{fig:BechmarkNindep} exhibits a clear improvement in success probability for intermediate interaction ranges. By contrast, Fig.~\ref{fig:BenchmarkNdep} shows a monotonic decrease in success probability. These results are consistent with the observations of Sec.~\ref{sec:numerical_results}. The numerical values illustrated in Figs.~\ref{fig:BenchmarkNdep}~and~\ref{fig:BechmarkNindep} are reported in Appendix~B.



\section{Conclusions\label{sec:conclusions}}
In this work, we have revisited the notion that Rydberg mediated neutral-atom quantum computing systems can increase connectivity and reduce gate synthesis overheads by resorting to ever higher principle quantum numbers, $n$. By investigating the trade-off between better connectivity versus reduced robustness to electric field noise at higher Rydberg levels, we find that as long as there is a finite, fluctuating background electric field, doing so ceases to be favourable once a sufficiently high principal quantum number is reached. This is because the gate error resulting from susceptibility to stray fields scales much more rapidly with respect to $n$ than any reduction in gate overheads due to a corresponding increase in connectivity.

The analysis presented here considered a model for noise that simplified the complicated noisy environment of current neutral-atom experiments down to two qualitatively distinct types of noise: one that depends on the range of Rydberg interactions and one that is independent of interaction range. This is a useful exercise because it makes clear that in most noise regimes, a preferred interaction range exists that best balances susceptibility to noise against gate synthesis overheads. This preferred interaction range is an important design parameter when engineering a quantum processor or compiler stack. Estimating it under realistic noise profiles can inform important design decisions (e.g.~how large can a monolithic quantum processor grow before technologies like interconnects become important~\cite{interconnects_2021}; or what assumptions vis-\`a-vis cluster sizes are safe to make when a compiler tries to map and resynthesize logical circuits onto hardware).

The precise interaction range that minimizes the overall error is of course specific to the noise model, atom species, and Rydberg level that we analyze. Including additional sources of noise that are independent of interaction range, such as finite atom temperature, would increase the preferred interaction range. Conversely, suppressing all range-independent noise sources would decrease it. For example, use of alkaline-earth atoms enables Rydberg transitions via a single-photon transition from a meta-stable groundstate, eliminating scattering from an intermediate energy level. Similarly, changing the characteristics of stray electric field fluctuations -- either by suppressing them overall or by changing their temporal or spectral profile so that they become more amenable to error-mitigation techniques -- would make a larger interaction range optimal.

It bears mentioning that Rydberg mediated gates are \textit{fundamentally} limited by the lifetime of the Rydberg level. While we did not explicitly model the effects of this finite lifetime in this work, we stress that this does \textit{not} change our conclusions -- that an interaction range ``sweet-spot'' exists under most noise regimes. Note that the Rydberg level lifetime grows (and the associated noise diminishes) with increasing principle quantum number, its spontaneous decay rate nominally scaling as $1/n^3$, with an additional contribution from black-body radiation induced stimulation emission that scales as $1/n^2$~\cite{Saffman_2010}. In the range of principle quantum numbers that we considered (i.e.~$57\leq n\leq110$), the Rydberg level lifetimes are calculated to be in the range $90-430\mu\text{s}$ ($200-1530\mu\text{s}$) at a temperature of $300\text{K}$ ($0\text{K}$). At small $n$'s these lifetimes are comparable in magnitude to the scattering times due to technical noise that we include in our noise model, whereas at larger $n$'s noise continues to be dominated by susceptibility to stray fields.

In sum, we believe our results and methodology for determining a preferred interaction range will be useful when designing applications and compilation techniques for neutral-atom computers. The flexibility of neutral-atom systems provides exciting avenues to further explore the role of connectivity on this QC platform. In particular, the adjustable atomic geometries enabled by optical tweezer traps offers an exciting opportunity to customize the QC topology informed by inherently preferred cluster sizes like those demonstrated here. Additionally, the promise of adiabatic entangling gates~\cite{Saffman_2020} and many atom gates~\cite{Dlaska_4body_2022} may offer a route to diminish the susceptibility to stray electric fields, while further reducing the gate overhead with larger interaction range.

\bibliography{bib}

\balancecolsandclearpage

\section*{Appendices}

\subsection{Loss scaling}

We wish to compute the probability of finding an excitation in a Rydberg level. Basic trigonometry yields:
\begin{align*}
p_{\text{loss}}=\left\langle \Pi_{11}\right\rangle  & =\frac{2\Omega^{2}}{\tilde{\Omega}(\delta)^{2}}\left(1-\cos\tilde{\Omega}(\delta)\tau\right)
\end{align*}
Note that if $\delta=0$, then $\cos \tilde{\Omega}(\delta)\tau=1$, and the loss probability vanishes, as in the case of an ideal LP gate. Now taking a Taylor expansion around $\tilde{\Omega}(\delta)\tau\approx 2\pi$, we have:
\begin{align*}
p_{\text{loss}}=\left\langle \Pi_{11}\right\rangle  & =\frac{\Omega^{2}}{\tilde{\Omega}(\delta)^{2}}\frac{\left(\omega\tau-2\pi\right)^{2}}{2}\\
 & =2\pi^{2}\frac{\Omega^{2}}{\tilde{\Omega}(\delta)^{2}}\left(\frac{\tilde{\Omega}(\delta)^{2}}{\tilde{\Omega}(0)^{2}}-2\frac{\tilde{\Omega}(\delta)}{\tilde{\Omega}(0)}+1\right)
\end{align*}

\vspace{5em}
Finally, noting that when $\delta/\Delta\approx0$:
\[
\frac{\tilde{\Omega}(\delta)}{\tilde{\Omega}(0)}\approx1+\frac{\Delta\delta}{\tilde{\Omega}(0)^{2}}+\frac{\Omega^{2}\delta^{2}}{\tilde{\Omega}(0)^{4}}...
\]
which yields:
\begin{align*}
p_{\text{loss}}=\left\langle \Pi_{11}\right\rangle  & \approx2\pi^{2}\Omega^{2}\frac{\delta^{2}\Delta^{2}}{\tilde{\Omega}(0)^{6}}
\end{align*}
Within the regime under consideration, $\delta\ll\Delta$, so the approximations we have made so far are justified. Ultimately, the only $n$-dependent term in the final expression is $\delta^{2}$. As we have seen $\delta\sim n^{7}$, which implies $p_{\text{loss}}\sim n^{14}$.

\onecolumngrid
\subsection{Benchmark results}

\begin{table}[h]
\begin{centering}
\begin{tabular}{|c|c|c|c|c|c|c|c|c|}
\hline
\multirow{2}{*}{Circuit} & \multicolumn{4}{c|}{$P_{\text{success}}$, only range-dependent noise} & \multicolumn{4}{c|}{$P_{\text{success}}$, with range-independent noise}\tabularnewline
\cline{2-9} \cline{3-9} \cline{4-9} \cline{5-9} \cline{6-9} \cline{7-9} \cline{8-9} \cline{9-9} 
 & $r_\text{max}=1$ & $r_\text{max}=\sqrt{2}$ & $r_\text{max}=2$ & $r_\text{max}=\sqrt{5}$ & $r_\text{max}=1$ & $r_\text{max}=2$ & $r_\text{max}=\sqrt{5}$ & $r_\text{max}=3$\tabularnewline
\hline 
\hline 
QFT (6-qb) & 99.98\% & 99.96\% & 99.85\% & 99.74\% & 96.22\% & 97.69\% & 98.02\% & 96.07\%\tabularnewline
\hline 
QFT (9-qb) & 99.96\% & 99.85\% & 99.53\% & 98.89\% & 89.59\% & 93.25\% & 94.35\% & 92.91\%\tabularnewline
\hline 
QFT (16-qb) & 99.85\% & 99.30\% & 97.93\% & 94.69\% & 60.50\% & 71.09\% & 74.74\% & 68.71\%\tabularnewline
\hline 
QPE (12-qb) & 99.96\% & 99.88\% & 99.58\% & 98.95\% & 88.98\% & 93.23\% & 94.23\% & 92.13\%\tabularnewline
\hline 
QPE (16-qb) & 99.92\% & 99.69\% & 99.04\% & 97.70\% & 78.67\% & 85.65\% & 87.55\% & 84.70\%\tabularnewline
\hline 
Adder (10-qb) & 99.98\% & 99.97\% & 99.80\% & 99.61\% & 95.61\% & 98.04\% & 97.87\% & 96.70\%\tabularnewline
\hline 
Adder (16-qb) & 99.98\% & 99.96\% & 99.69\% & 99.33\% & 94.85\% & 96.89\% & 97.02\% & 94.34\%\tabularnewline
\hline 
\end{tabular}
\par\end{centering}
\caption{Table of success probabilities, $P_{\text{success}}$, for a variety of quantum circuits when executed with and without noise that is independent of the interaction range. For each indicated maximum interaction radius, $r_\text{max}$, the corresponding principle quantum number $n$ used to realize an LP gate is chosen in a ``graded'' fashion, as discussed in Sec.~\ref{subsec:N-dependent-results}.}
\end{table}

\end{document}